\documentclass[aps,prl,twocolumn,showpacs]{revtex4}
\usepackage{graphicx}
\usepackage{amsmath}

\begin{document}

\title{Polar charge and orbital order in {\it 2H}-TaS$_2$}

\author{Jasper van Wezel}
  \affiliation{Materials Science Division, Argonne National Laboratory, Argonne, IL 60439, USA}

\begin{abstract}
It was recently discovered that in spite of the scalar nature of its order parameter, the charge order in {\it 1T}-TiSe$_2$ can be chiral. This is made possible by the emergence of orbital order in conjunction with the charge density modulations. Here we show that a closely related charge and orbital ordered state arises in {\it 2H}-TaS$_2$. In both materials, the microscopic mechanism driving the transition is based on the interaction between three differently polarized displacement waves. The relative phase shifts between these waves lead both to the formation of orbital order and to the breakdown of inversion symmetry. In contrast to {\it 1T}-TiSe$_2$ however, the presence of a mirror plane in the lattice of {\it 2H}-TaS$_2$ prevents the distorted structure in this material from being chiral, and a polar charge and orbital ordered state arises instead. It is stressed that bulk experiments are indispensable in differentiating between the chiral and polar phases.
\end{abstract}


\maketitle

\section{Introduction}
The emergence of charge density wave (CDW) order in a material implies modulations in both its electronic density and, because of the non-zero electron-phonon coupling, the equilibrium positions of its atomic lattice. While the electronic density modulations already break the translational symmetry of the underlying lattice, and profoundly affect both the transport and equilibrium properties of the material \cite{Peierls55:book,Gruner88}, the ionic displacements may in certain materials conspire to break additional symmetries, and give rise to supplemental ordering. The breakdown of rotational symmetry in dichalcogenides for example, can give rise to a structural transition within the charge ordered phase \cite{Littlewood82}, while the broken inversion symmetry in SnTe allows the appearance of ferroelectricity in conjunction with the CDW in that material \cite{Littlewood84}.

Only recently, it was discovered that it is possible for the atomic displacements accompanying charge order to break the inversion symmetry of the atomic lattice in a chiral way \cite{Ishioka10,JvW:Physics11}. How such a helical configuration of charge density can be formed is not {\it a priori} obvious, because CDW materials, unlike spin density waves, have a scalar order parameter which cannot straightforwardly be made chiral. This apparent paradox was resolved by interpreting the observed chiral CDW phase in {\it 1T}-TiSe$_2$ as the result of simultaneous orbital and charge order \cite{JvW:arXiv11}. Since the prerequisites for the formation of such a chiral charge and orbital ordered phase are rather generic, it may be expected that a chiral CDW can be identified in a broad class of charge ordered materials \cite{JvW:ecrys11}. 
\begin{figure}[t]
\centerline{{\includegraphics[width=0.85 \columnwidth]{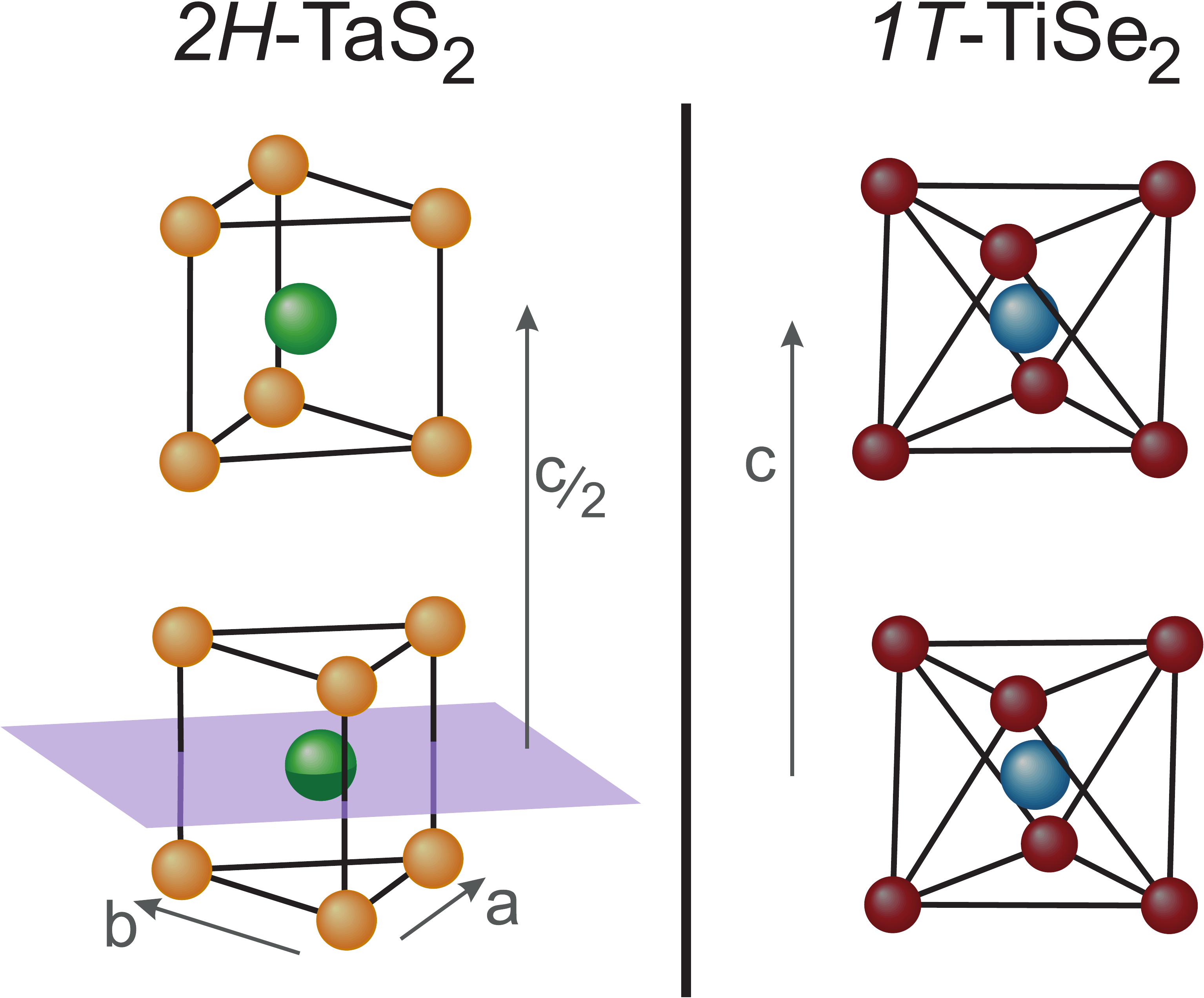}}}
\caption{A comparison between the lattice structures of {\it 2H}-TaS$_2$ and {\it 1T}-TiSe$_2$. {\bf Left}: a single unit cell of {\it 2H}-TaS$_2$. The trigonal prismatic coordination of the Ta atoms is invariant under reflection in the indicated mirror plane. This symmetry is not broken by the formation of charge order. {\bf Right}: two unit cells of {\it 1T}-TiSe$_2$. Notice that, owing to the octahedral coordination of the Ti atoms, the mirror plane is absent in this structure.}
\label{lattice}
\end{figure}

Based on an analysis of scanning tunneling microscope (STM) images of its surface, it was recently suggested that the CDW in {\it 2H}-TaS$_2$ closely resembles the charge order seen in {\it 1T}-TiSe$2$, and in particular that it is also chiral \cite{Guillamon11}. This suggestion is especially interesting because at low temperatures, the charge density modulations in {\it 2H}-TaS$_2$ are known to coexist with a superconducting phase. The emergence of superconductivity from a state with chiral electronic structure could affect both the spin structure of its Cooper pairs and the symmetry of the gap, although no such effects were observed in this case \cite{Guillamon11}.
\begin{figure*}[t]
\centerline{{\includegraphics[width=0.9 \textwidth]{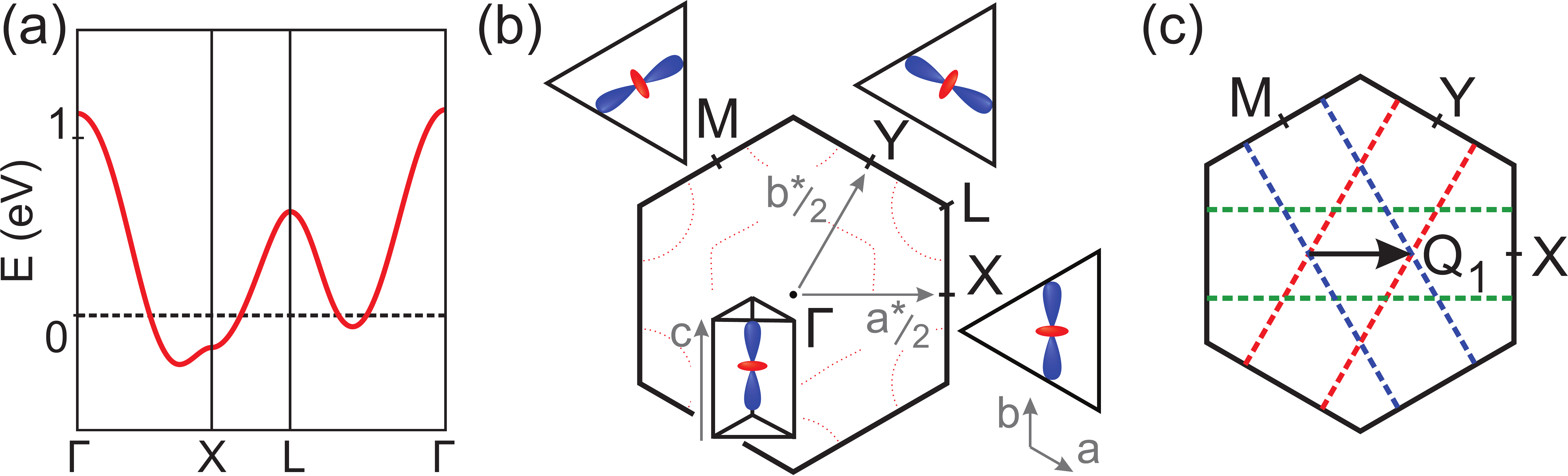}}}
\caption{The electronic structure of {\it 2H}-TaS$_2$. {\bf (a)} The electronic band structure close to the Fermi energy (adapted from \onlinecite{Whangbo92}). Only the lowest lying band derived from Ta-$d$ electrons, and none of the S-derived bands, cross the Fermi level. {\bf (b)} The orbital character of the electrons in the lowest Ta-$d$ band. The real space orbital shape of electrons with a given momentum is shown for the electronic states at high symmetry points in the first Brillouin zone. At the zone centre, the electrons are aligned perpendicular to the Ta planes, while at the zone edges the orbitals lie within that plane. The dotted line gives a schematic indication of the position of the Fermi surface in the first Brillouin zone. {\bf (c)} The Fermi surface sheets obtained from the model of non-interacting one-dimensional chains. The projection of the first Brillouin zone to the $a^*$-$b^*$-plane shows flat sheets of Fermi surfaces extending along the $c^*$ direction as straight dashed lines. Each pair of sheets corresponds to a particular orientation of Ta-$d$ orbitals. Two pairs of sheets can be simultaneously connected by a single nesting vector, as shown for $\vec{Q}_1$.}
\label{BZ}
\end{figure*}

Here we point out that in spite of the apparently chiral structure observed in STM experiments, the lattice symmetry of {\it 2H}-TaS$_2$ does in fact not allow the formation of a helical state. It is shown that as in {\it 1T}-TiSe$_2$, the charge order in this compound arises from a microscopic mechanism which involves orbital ordering as well as a modulation of the charge density. Unlike the {\it 1T} structure however, the lattice of the {\it 2H} material has a mirror symmetry, which prevents its CDW from becoming chiral (see Fig. \ref{lattice}). Using a Landau order parameter analysis we show how instead a polar orbital and charge ordered state arises in {\it 2H}-TaS$_2$. This polar state looks identical to the chiral state in {\it 1T}-TiSe$_2$ if only the topmost atomic layers are considered, as observed in the recent STM experiments. Bulk measurements which may be employed to confirm the presence of a polar charge and orbital ordered state are proposed.

\section{One-dimensional chains}
The hexagonally ordered sheets of Tantalum atoms in the quasi two-dimensional, layered transition metal dichalcogenide {\it 2H}-TaS$_2$ are sandwiched between sheets of Sulphur atoms. The local coordination of the Ta atoms is trigonal prismatic, as  indicated in Fig. \ref{lattice}, resulting in a crystal field splitting of the Ta-$d$ orbitals \cite{Huisman71}. It is known from band structure calculations that only the lowest Ta-$d$ band, and none of the Se-$p$ bands, cross the Fermi energy, as shown in Fig. \ref{BZ}a  \cite{Mattheiss73,Whangbo92,Inglesfield00}. It may thus be expected that the CDW in {\it 2H}-TaS$_2$ results predominantly from the hybridization of its metal atoms \cite{Whangbo92}.

The orbital character of the $d$-bands can be obtained from a tight binding fit to the band structure. At the centre of the Brillouin zone, the electrons in the band closest to the Fermi energy are purely of the $d_{z^2}$ type, with the $z$ direction oriented along the crystallographic $c$-axis \cite{Mattheiss73,Whangbo92}. Going towards the edges of the Brillouin zone, the character of the electrons changes, until precisely at the edge (the X-point in Fig. \ref{BZ}b), the orbital is again of the $d_{z^2}$ type, but now with its the quantization axis along the crystallographic $b$-axis. Going around the edge of the Brillouin zone, the shape of the orbital remains the same, but its orientation is rotated so that at the Y and M points in Fig. \ref{BZ}b they are aligned with the crystallographic $a$ and $-a-b$ axes respectively. All of these orbital wave functions can be written as different linear combinations of the $d_{z^2}$, $d_{x^2}$ and $d_{y^2}$ orbitals, defined with respect to the $c$-axis.

The Fermi surface of {\it 2H}-TaS$_2$ lies approximately two thirds of the way between the centre of the Brillouin zone and its edges. As a first order approximation, we can thus describe the electrons at the Fermi surface using the three in-plane $d_{z^2}$-like orbitals which become exact at the zone boundaries. These orbitals overlap strongly along one-dimensional chains in the crystal lattice, while the exchange integrals with neighboring chains are small. Based on this observation, it has been suggested that the charge density wave formation in this material may be understood as the result of the `hidden nesting' of non-interacting, one-dimensional chains of Ta orbitals \cite{Whangbo92}, in close analogy to the mechanism underlying for example the formation of the charge ordered structure in elemental Tellurium \cite{Fukutome84,JvW:Physics11}.

The four different $d_{z^2}$-like orbitals are not mutually orthogonal, and form an overcomplete basis for the description of the lowest $d$-band. The approximate description of {\it 2H}-TaS$_2$ as a system of non-interacting one-dimensional chains can therefore not be tenable throughout the first Brillouin zone. In particular, we know that the interactions between in-plane orbitals, which are strictly zero at the points X, Y, and M, must become strong at the Brillouin zone centre. We will nevertheless neglect these interactions here, because the tight binding fit of the band structure shows that at the zone centre, the lowest band lies far above the Fermi energy. As long as we only consider low energy processes which occur close to the Fermi level, the description in terms of non-interacting chains will be a good approximation. We thus consider the Hamiltonian
\begin{align}
\hat{H} =2 t \sum_k & \left\{ \cos(\vec{k} \cdot \vec{a}) \hat{a}^{\dagger}_k \hat{a}^{\phantom \dagger}_k  + \cos(\vec{k} \cdot \vec{b}) \hat{b}^{\dagger}_k \hat{b}^{\phantom \dagger}_k \right. \notag \\
& \ \ \ + \left. \cos(\vec{k} \cdot \vec{d}) \hat{d}^{\dagger}_k \hat{d}^{\phantom \dagger}_k \right\}.
\label{H_1D}
\end{align}
Here $\vec{a}$, $\vec{b}$ and $\vec{d}=-\vec{a}-\vec{b}$ are unit cell vectors of the {\it 2H} crystal structure (see Figs. \ref{lattice} and \ref{BZ}b), and the three different fermionic operators create electrons in the three different in-plane $d_{z^2}$-like states.

The three one-dimensional chains share one electron per unit cell, resulting in three sets of flat Fermi surface sheets, as indicated in Fig. \ref{BZ}c. The perfect nesting of these sheets gives rise to three concurrent CDW instabilities. Choosing the propagation vectors of these instabilities so that each charge density wave simultaneously modulates the electron distributions of two different sets of electrons, allows the system to maximize its energy gain \cite{Whangbo91}. We thus recover the three experimentally observed CDW vectors $Q_1=a^* / 3$, $Q_2=(-a^*+b^*) / 3$, and $Q_3=-b^* / 3$ \cite{Wilson75,Moncton77,Brouwer80}.

The atomic displacement waves accompanying the charge order will share its propagation vectors, but their polarizations depend on the anisotropy of the electron-lattice coupling as determined by the orbital structure \cite{JvW:arXiv11}. Since each species of in-plane orbital is arranged along one-dimensional chains with negligible electronic overlap between them, their individual displacements can only be directed along the chains. Each nesting vector connects two types of Fermi surface sheets, belonging to two differently oriented orbitals. The overall displacement imposed by the wave propagating along a given nesting vector is thus the linear superposition of the displacements induced in the two orbital sectors affected by it \cite{Fukutome84,JvW:arXiv11}. Combining all CDW components, we then find the total displacement pattern
\begin{align}
\vec{u}(\vec{r}) \propto & [ \vec{d} -\vec{a} ] \sin( \vec{Q}_1 \cdot \vec{r} ) +  [ \vec{a} -\vec{b} ] \sin( \vec{Q}_2 \cdot \vec{r} ) \notag \\
 & \  + [ \vec{b} -\vec{d} ] \sin( \vec{Q}_3 \cdot \vec{r} ).
\label{3q}
\end{align}
Notice that the present analysis does not uniquely determine the relative signs of the polarization components. With the signs as in Eq. \eqref{3q}  however, we precisely reproduce the lattice structure that had been suggested for the CDW phase of {\it 2H}-TaS$_2$ before the possibility of a chiral phase became apparent \cite{Wilson75,Moncton77,Brouwer80}.

\section{Relative phases}
To describe the microscopic interactions between the electronic density modulations and the lattice which determine the relative phases of the displacement wave components, we construct a Ginzburg-Landau theory. The order parameter consists of the modulation $\alpha(\vec{r})$ of the average charge density, which can be written as a sum of three complex components $\psi_j= \psi_0 e^{i(\vec{Q}_j\cdot \vec{r} + \varphi_j)}$. The amplitudes of all components are assumed to be equal and the propagation vectors $\vec{Q}_j$ are the CDW nesting vectors. The phases $\varphi_j$ are to be determined from the minimization of the free energy, which is given by \cite{McMillan75}:
\begin{align}
F = \int d\vec{x} & \left\{ \ a\alpha^2 + b \alpha^3 + c \alpha^4 + d \alpha^6 \right. \notag \\
&+ \left. f \left[ |\psi_1 \psi_2 |^2 + |\psi_2 \psi_3 |^2 + |\psi_3 \psi_1 |^2 \right] \right\}.
\label{F}
\end{align}
The powers of $\alpha$ represent the terms allowed by symmetry in the expansion of both the electronic Coulomb energy and the elastic energy cost of deforming the lattice. The cross terms in the last line signify the competition between different CDW components over the available Fermi surface.

To  take into account also the Umklapp processes associated with the presence of a discrete lattice, we expand the coefficients of the expression for the free energy such that they reflect the lattice symmetry \cite{McMillan75}.
\begin{align}
a = a_0 + a_1 \sum_i e^{i\vec{G}_i \cdot \vec{r}} + \gamma a_1 \sum_{i,i'} e^{i\vec{G}_i \cdot (\vec{r}+\vec{R}_{i'} ) } + .. 
\end{align}
Here $\vec{G}_i$ are the shortest reciprocal lattice vectors and $\vec{R}_{i}$ connect a Tantalum site to its Sulphur neighbors in the same unit cell. Because $a_1$ arises from Umklapp effects, it is proportional to the electron-phonon coupling. The factor $\gamma$ reflects the difference in strength of the electron-phonon coupling on Ta and S sites. From the tight-binding fit to the band structure, we know that the distortions in {\it 2H}-TaS$_2$ result from the hybridization of Ta-$d$ orbitals, with the Sulphur atoms playing only a secondary role. We therefore assume $\gamma$ to be small, and set it to zero from here on. 

The free energy of Eq. \eqref{F} can be evaluated directly, and its phase dependent part is given by
\begin{align}
F & = \frac{3}{2} \psi_0^3 b_0^{\phantom 3} \cos( \varphi_1 + \varphi_2 + \varphi_3) \notag \\
& + \frac{1}{4} \psi_0^3 b_1^{\phantom 3} \sum_j \cos(3 \varphi_j) \notag \\
& + \frac{3}{2} \psi_0^4 c_1^{\phantom 4} \sum_j \cos(2 \varphi_j - \varphi_{j+1} - \varphi_{j-1}) \notag \\
& + \frac{5}{8} \psi_0^6 d_0^{\phantom 6} \sum_j \cos(3 \varphi_j - 3 \varphi_{j+1}),
\label{Fphase}
\end{align}
where we took into account terms up to sixth order. The coefficients $b_1$ and $c_1$ arise from the coupling of the electronic charge modulation to the ionic lattice. To favor bonding between Ta atoms, $b_1$ should be chosen positive, and $c_1$ negative. The minimum energy solution up to fourth order is then the normal CDW with $\varphi_1=\varphi_2=\varphi_3=\pi / 3$, which corresponds to the non-chiral structure mentioned before. Including a finite, non-zero value for $d_0$ however, changes the solution qualitatively. The phases of the different charge density wave components can no longer all be equal, and the corresponding atomic displacements break the inversion symmetry of the lattice. In the extreme case with $d_0 \gg c_1$,  the stable solution obeys $\varphi_1 = \pi/3$, $\varphi_2 = \varphi_1 \pm 2 \pi /9$ and $\varphi_3 = \varphi_1 \pm 4 \pi/9$, while for intermediate values of $d_0$, the phase differences interpolate smoothly between the two extreme solutions.
 
Even though $d_0$ enters the expansion of the free energy only at sixth order, it arises directly from the electronic Coulomb interactions, while the fourth-order coefficient $c_1$ corresponds to an Umklapp effect, and is proportional to the electron-phonon coupling on the Ta sites. The ratio $d_0 / c_1$ is therefore not in general negligibly small, and the interplay between Coulomb interaction and electron-lattice coupling may give rise to the formation of non-zero phase differences. The pattern observed in recent STM experiments on {\it 2H}-TaS$_2$, which was suggested to be chiral \cite{Guillamon11}, corresponds to such a regime, in which the sixth order term does not dominate the expression of Eq. \eqref{Fphase}, but also cannot be neglected.

\section{The polar CDW}
\begin{figure}[t]
\centerline{{\includegraphics[width=0.9 \columnwidth]{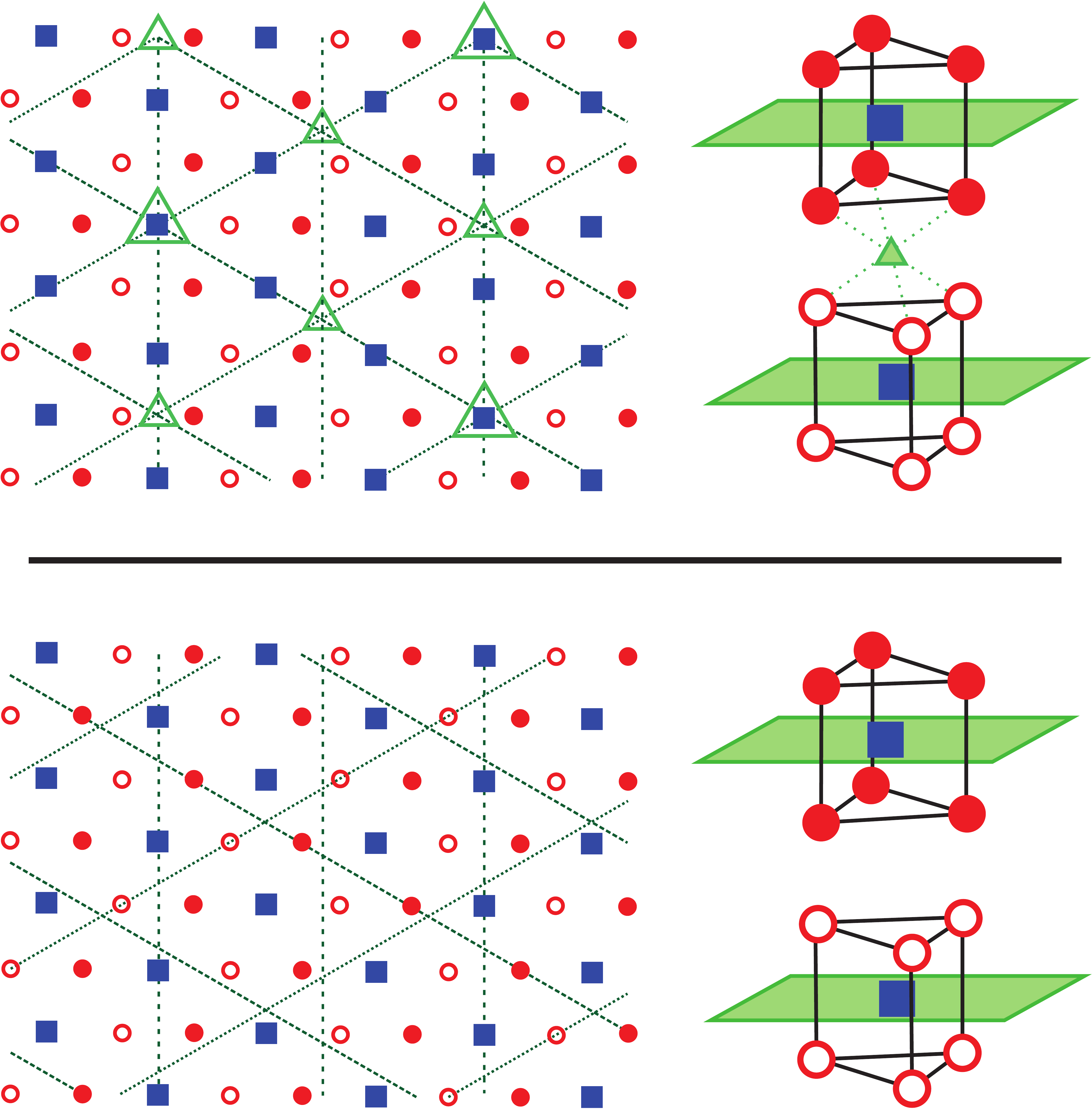}}}
\caption{The non-polar and polar charge density waves. {\bf Top}: The extrema of the three components of the charge density modulation $\alpha=\sum_j \psi_0 \cos(\vec{Q}_j\cdot \vec{r} + \varphi_j)$, in the absence of any relative phase differences, are drawn as dashed lines in the planar projection on the left. Centres of inversion, indicated by triangles, exists at the points in the plane where three extrema meet. The position of one of the inversion centres within the three-dimensional atomic structure is shown schematically on the right, along with two horizontal mirror planes. {\bf Bottom}: For non-zero relative phase differences, the extrema of two charge density wave components are shifted. The resulting structure no longer has any centre of inversion. The horizontal mirror planes indicated in the three dimensional structure however, are unaffected by the relative phase shifts. The atomic lattice in this state thus breaks inversion symmetry, but retains its mirror symmetries, so that it is polar rather than chiral.}
\label{polar}
\end{figure}

The effects of the absence or presence of relative phase differences between the charge density wave components are schematically indicated in Fig. \ref{polar}. In the former case, the extrema of all three components of the charge density wave conicide on or in between Ta sites. The resulting atomic displacements do not break the inversion symmetry operations centered at the same positions, and the distorted lattice structure concurs with the previously reported $3 \times 3$ superstructure \cite{Wilson75,Moncton77,Brouwer80}.  If the relative phases become non-zero however, the extrema of two components start to shift. The extreme situation with $d_0$ dominating Eq. \eqref{Fphase}, is shown in the bottom of Fig. \ref{polar}. In practice, this extreme situation will not be reached, and the actual shifts of the charge density wave extrema in {\it 2H}-TaS$_2$ will be smaller. Regardless of the size of the relative phase shifts however, the inversion centres of the original lattice will no longer coincide with the extrema of all three charge density wave components, and the inversion symmetry of the lattice is broken.

The lattice structure of the {\it 2H} polytype, as shown in Figs. \ref{lattice} and \ref{polar}, consists of two TaS$_2$ sandwich layers per unit cell, and the top and bottom Sulphur atoms within a single sandwich are affected by the charge density modulations in the same way. In contrast to {\it 1T}-TiSe$_2$, the central Ta atoms therefore always define a mirror plane, even in the presence of non-zero relative phases. Since it is not possible to define left or right handed patterns in the presence of this mirror symmetry, the distorted lattice structure is not chiral. The modulations do however break the inversion symmetry of the lattice, so that the final lattice structure is found to be polar. 

The polar and the non-polar phases shown in Fig. \ref{polar} both have $3 \times 3$ superstructures defined by the same set of CDW propagation vectors, and differ only in their values for the relative phases between the charge density wave components. Both phases therefore agree with previously reported neutron scattering and X-ray diffraction measurements of the electronic superstructure \cite{Wilson75,Moncton77,Brouwer80}. 

The atomic displacements of only the top two atomic layers do not show the presence of the mirror plane or the absence of chirality, so that their Fourier transform will look identical in character to the patterns observed in a chiral state. The presence of relative phase differences leads to a displacement of the extrema of two CDW components away from the central Ta atoms, thus giving rise to the observed variations in electronic density on these atoms. The similarity of the recent STM images of the surface of {\it 2H}-TaS$_2$ to those obtained on chiral {\it 1T}-TiSe$_2$ clearly indicates that both materials harbor a three-component charge density wave, and that in both materials the phase differences between the components can be non-zero \cite{Guillamon11,Ishioka10,JvW:arXiv11}. Whereas in the case of {\it 1T}-TiSe$_2$ those phase differences imply a chiral lattice structure, in {\it 2H}-TaS$_2$ the displacements form a polar pattern. 

Experimental verification of the proposed polar lattice structure requires bulk probes. The reduction of the space group of {\it 2H}-TaS$_2$ from P6$_3$/mmc (no. 194) to Pm (no. 6) as the polar phase is entered should be observable in for example X-ray diffraction experiments. Indirect evidence of the absence of chirality may also be gained from optical reflectometry. The polarization dependence of reflected light was shown to be two-fold symmetric in the chiral phase of {\it 1T}-TiSe$_2$, in contrast to the three-fold symmetry expected for a non-chiral three-component charge density wave \cite{Ishioka10}. 

\begin{figure}[t]
\centerline{{\includegraphics[width=0.7 \columnwidth]{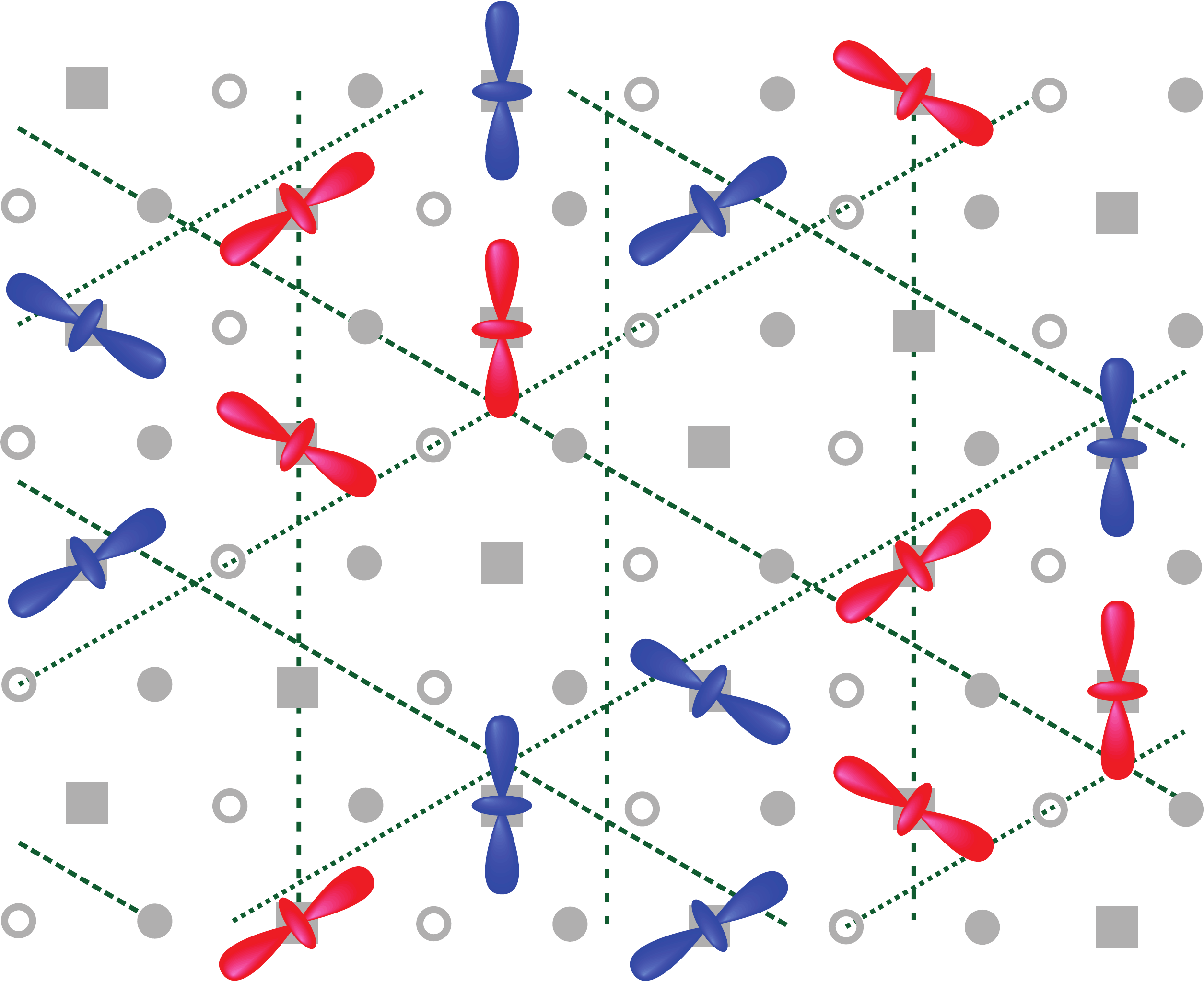}}}
\caption{The orbital order in the polar phase of {\it 2H}-TaS$_2$. Only those Ta-$d$ orbitals whose occupation varies most significantly as the system goes from the uniform to the polar phase, are indicated in the planar projection of the atomic lattice. Red orbitals identify a decrease in occupation, while blue orbitals represent an increase. As in fig. \ref{polar}, the dashed lines correspond to the extrema of the three charge density wave components. Notice that the orbital order directly evidences the breakdown of both rotational symmetry and inversion symmetry in the polar phase, but does not affect the planar mirror symmetry.}
\label{orbital}
\end{figure}

Finally, like in {\it 1T}-TiSe$_2$, the shifting of the CDW components due to the non-zero phase differences also automatically implies the formation of orbital order in the system. The polarizations of the displacement wave components in Eq. \eqref{3q} each originate in a combination of two specific orbital sectors. The relative differences between the amplitudes of the three charge density wave components on a given Ta site therefore also correspond to relative differences between the electronic occupations of the three in-plane Ta-$d$ orbitals. In the polar phase, the variations in orbital occupation are schematically shown in Fig. \ref{orbital}, in which only the orbitals most significantly affected by the charge modulations are drawn. Notice that the variations in occupation between different sites define an orbital density wave, rather than an arrangements of completely filled and empty orbitals. The presence of this type of orbital order in {\it 2H}-TaS$_2$ is a direct consequence of the presence of the relative phase differences which define the polar charge density wave.
%

\section{Conclusions}
In summary, we have shown that the lattice structure in the charge ordered phase of {\it 2H}-TaS$_2$ may be understood as resulting from the interaction between three differently polarized displacement waves. The longitudinal polarizations of these waves are due to the relative orientations of the electronic orbitals involved in the CDW formation. As in the similar compound {\it 1T}-TiSe$_2$, the interplay between the local Coulomb interaction and the electron-lattice coupling gives rise to the presence of non-zero relative phase differences between the components of the charge density modulation, which break the inversion symmetry of the overall structure. Unlike the case of {\it 1T}-TiSe$_2$ however, the distortions in the lattice of {\it 2H}-TaS$_2$ yield a polar, rather than a chiral, phase. The emergence of chirality in this material is forestalled by the presence of a mirror plane in its lattice, which cannot be destroyed by the phase shifts of the three CDW components. Bulk experiments are indispensable in identifying whether a given material is chiral or polar, because surface probes are not sensitive to the existence of this mirror symmetry.

In spite of the different character of their final atomic configurations, the charge density waves in {\it 1T}-TiSe$_2$ and {\it 2H}-TaS$_2$ are closely related. Both of them arise from similar `hidden' one-dimensional chains of orbitals, which give rise to differently polarized components of a displacement wave. The central role played by the electronic orbitals in defining these polarizations, is reflected in the fact that the formation of both this type of polar and of chiral charge order are necessarily accompanied by the emergence of orbital order. Finally, the phase shifts which give rise to the breakdown of inversion symmetry in both compounds arise from a balance between similar terms in the Landau expansion of their free energies. 

These similarities, and the generic nature of the interactions underlying them, strongly suggest that orbital order and the loss of inversion symmetry may be general characteristics of a much broader group of materials with multiple charge density waves.

\subsection{Acknowledgements}
This research was supported by the US DOE, Office of Science, under Contract No. DE-AC02-06CH11357.


\end{document}